\documentclass[twocolumn,showpacs,preprintnumbers,prb,floatfix,amsmath,amssymb,superscriptaddress]{revtex4-1}
\usepackage{newlfont,epsfig}
\usepackage{graphicx}
\usepackage{epstopdf}
\usepackage[bf,SL,BF]{subfigure}

\begin{document}

\title{Exchange interaction and tunneling induced transparency in coupled quantum dots}
\author{H. S. Borges}
\affiliation{Instituto de Fisica, Universidade Federal de
Uberl\^andia, 38400-902, Uberl\^andia-MG, Brazil}
\affiliation{Department of Physics and Astronomy, and Nanoscale and Quantum Phenomena Institute, Ohio University,
Athens, Ohio 45701-2979, USA}
\author{A. M. Alcalde}
\affiliation{Instituto de Fisica, Universidade Federal de
Uberl\^andia, 38400-902, Uberl\^andia-MG, Brazil}
\author{Sergio E. Ulloa}
\affiliation{Department of Physics and Astronomy, and Nanoscale and Quantum Phenomena Institute, Ohio University,
Athens, Ohio 45701-2979, USA}

\begin{abstract}
We investigate the optical response of quantum dot molecules coherently driven by polarized laser light.  Our description includes the splitting in excitonic levels caused by 
isotropic and anisotropic exchange interactions. We consider interdot transitions mediated by hole tunneling between states with the same total angular 
momentum and between bright and dark exciton states, as allowed by spin-flip hopping between the dots in the molecule. Using realistic experimental 
parameters we demonstrate that the excitonic states coupled by tunneling exhibit a rich and controllable optical response. We show that through the 
appropriate control of an external electric field and light polarization, the tunneling coupling establishes an efficient destructive quantum interference 
path that creates a transparency window in the absorption spectra, whenever states of appropriate symmetry are mixed by the carrier tunneling. We 
explore the relevant parameter space that allows probing this phenomenon in experiments.  Controlled variation of applied field and laser detuning 
would allow the optical characterization of spin-preserving and spin-flip hopping amplitudes in such systems, by measuring the width of the 
tunneling-induced transparency windows. 

\end{abstract}
 \maketitle

\section{Introduction}
\label{sec:intro}

It is well established that interference between different excitation paths can control the optical response of a quantum system.  This leads, for example, to 
the suppression of light absorption when interference between different channels is destructive. Electromagnetically induced transparency, 
where the absorption of a probe laser is suppressed by the presence of a strong coupling laser, has been intensively investigated and demonstrated in 
atomic and semiconductor systems.\cite{Gerardot09, Celso10, Imamoglu12, Ma13}

Stacked semiconductor quantum dots coupled by tunneling, also known as quantum dot molecules (QDMs), are systems where it is possible to create a 
multilevel manifold of excitonic states which may considerably enrich quantum interference phenomena with possible applications in quantum
information systems.\cite{QDMbook}
The strong 
confinement of exciton states as well as the ability to control their coupling through external electric fields enables the investigation of optical 
interference induced by interdot tunneling and applied optical fields. Thus, under certain conditions, the tunneling coupling establishes an efficient destructive 
quantum interference path, which creates a sharp and switchable transmission region in the absorption spectra. This effect is analogous to the electromagnetic induced 
transparency, where the role of the optical pump field is played here by the strong tunneling coupling between the quantum dots. Such {\em tunneling 
induced transparency} has been recently investigated in coupled quantum dots in which the conduction band levels are brought into resonance by a 
gate voltage.\cite{Yu13, Zhu06}  One can then say that the transparency is caused by the presence of the coherent electron tunneling between dots. 

Several groups have demonstrated the tunneling of electrons and holes between coupled quantum dots.\cite{Stinaff06, Gammon06, Dotty10}
However, systems based in hole state tunneling may be superior in
the context of quantum information applications. For example, the weak 
hyperfine coupling of holes with nuclear spins make the hole spin more robust to dephasing effects induced by the nuclear spins.\cite{Gammon11}
Although our model is rather flexible, we focus our study on QDM structures where hole tunneling is dominant by design.

The presence of strain fields during the process of formation of the QD nanostructure as well as the shape anisotropy exhibited by self-assembled dots in 
III-V materials causes the excitonic spectrum to be strongly influenced by the exchange interaction. This arises between electron and hole in the
exciton states due to their
Coulomb interaction and involves their spins in the resulting level manifold.\cite{Bayer02, Muller04, Villas07} 
The asymmetry of the confinement potential in the QD and its impact in the Coulomb integrals that mediate the exchange interaction produce 
splitting of neutral exciton levels, leading to the well-known exciton ``fine structure". \cite{Bayer02,Gammon96} 
The response of such quantum system to polarized laser light is then strongly affected by the exchange 
interaction. Tunneling induced transparency in this system, when exchange interactions are explicitly taken into account, has not 
been explored yet.  As we will show below, this effect results in very interesting optical behavior of the quantum dot molecule system,
including transparency windows and absorption peaks that are sensitive to electric field and optical detuning.

In this work we investigate the optical linear susceptibility and absorption of a weak probe laser in a QDM system, as one varies an applied electric field and the controlled detuning of the laser near the desired excitonic states.  We consider the exciton fine structure splitting caused 
by isotropic and anisotropic 
exchange interactions. We solve the Liouville-von Neumman-Lindblad equation numerically in the Markovian approximation, where
the spontaneous exciton 
decay is considered the main decoherence mechanism. Our results show that the optical response is indeed strongly affected by the electron-hole exchange interaction and by appropriately tuning an applied electric field.  The hole tunneling coupling can induce a transparency window 
on the otherwise absorption spectrum of the system. The width of this transparency window is directly proportional to the inter dot hopping amplitude.
As such, optical measurements under controlled conditions would provide one with information on the effective interdot coupling, as we will explain.  

The paper is organized as follows: in Sec.\ \ref{sec:theory} we discuss the physical system and model. Section \ref{sec:chi} is devoted to 
analyzing the optical susceptibility considering the effects of tunneling, decoherence mechanism and exchange interactions.  Section \ref{sec:concl} 
presents concluding remarks.

\section{Physical system and model}
\label{sec:theory}

We consider a III-V QDM with a typical structural asymmetry, such that the dimensions of the bottom dot (especially its height)
are larger than on the top dot. \cite{Stinaff06}
Resonant optical excitation typically produces spatially direct excitons (where the electron and hole reside in the same dot), while
the electron or hole tunneling enables the formation of spatially indirect excitons. An electric field $F$ applied along 
the growth direction induces a relative Stark shift of the levels in both quantum dots. While direct excitons are weakly shifted, indirect excitons show a 
strong Stark blueshift with field. 
As the field varies, anticrossings between direct and  indirect excitons are produced by tunneling at well
defined values of the field. In our model we consider that the electron is confined in the bottom dot, and that the valence band levels of top and bottom dots 
are nearly resonant, favoring hole tunneling between the dots, as realized in different experiments.\cite{Gammon06}

We model the dynamics of all energetically relevant spatially direct and indirect exciton states in the double QD coupled by hole tunneling. 
We consider heavy hole (hh) states $|j_{hh},m_{hh} \rangle = |3/2, \pm 3/2 \rangle$ and conduction electrons $|j_{e},m_{e} \rangle = |1/2, \pm 1/2 
\rangle$, as the relevant manifold in our problem. As a consequence, the $z$-component of the exciton angular momentum can assume the values $S_z = 
m_{hh}+m_{e}=\pm 1$ (bright excitons) and  $\pm 2$ (dark excitons). The hole tunneling, moreover, can take place either changing or conserving its 
angular momentum projection. The spin relaxation can occur mainly due to the hyperfine interaction between the carrier spin 
and the nuclear spins in the system or as a consequence of the spin-orbit interaction. \cite{Chekhovich13} 
The hole tunneling rates are denoted by: $T_h^f$ ($T_h$) for flipping (conserving) intrinsic angular momentum 
projection. In this way, the dark states, associated with parallel spin configurations, and 
essentially disconnected from the dynamics via optical excitations become 
now accessible and play in fact an important role, as we will see.  
    
In quantum dots, the electron-hole exchange interaction (EHEI) produces significant features in the QD excitonic spectra, such as the splitting between 
bright exciton ($S_z=\pm 1$) and dark exciton ($S_z=\pm2$) states. Also, the anisotropy of the confinement potential in the QD plane enables the 
splitting of both bright and dark exciton doublets.  This anisotropic exchange interaction plays an essential role in optical excitations with polarized light. \cite{Gammon96, Economou08, Villas07} As EHEI is directly 
related to the electron-hole wave function overlap, the resulting exchange splitting for direct excitons is larger than for indirect exciton states. 
Thus, we model 
the QDM using four direct states $|d_{\pm 1}\rangle$, $|d_{\pm 2}\rangle$ and four indirect states $|i_{\pm 1}\rangle$, $|i_{\pm 2}\rangle$. The exciton 
vacuum is labeled as $|0\rangle$.   

\begin{figure}[htb]
\centering
\includegraphics[scale=0.8]{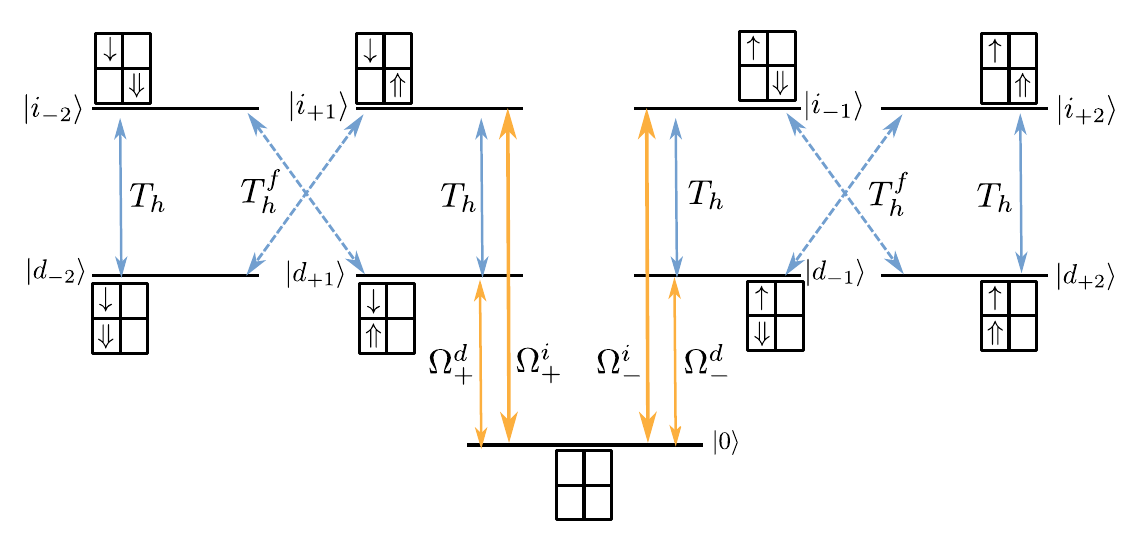}
\caption{(Color online) Schematic representation of exciton levels including the relevant couplings given in hamiltonian (\ref{eq:1}), where $\Uparrow$ and
$\Downarrow$ represent holes and $\uparrow$ and $\downarrow$ the electron states, while left/right boxes denote different dots in the
QDM.}\label{Fig1}
\end{figure}

The hamiltonian that describes the system in the rotating-wave approximation is given by (see Fig.\ \ref{Fig1})
\begin{equation}
\label{eq:1}
H=H_0+H_p+H_{T_h}+H_{T_h^f}+H_{\mathrm exch},
\end{equation}
with
\begin{eqnarray*}
H_0=\sum_{\substack{S_z=\pm2,\pm1}}\delta_d |d_{S_z}\rangle \langle d_{S_z}|+\left(\delta_i-\Delta_F\right)|i_{S_z}\rangle \langle i_{S_z}|+h.c.
\end{eqnarray*}
\begin{eqnarray*}
H_p=\Omega_{\pm}^d|d_{\pm1}\rangle \langle 0|+\Omega_{\pm}^i|i_{\pm1}\rangle \langle 0|+h.c.
\end{eqnarray*}
\begin{eqnarray*}
H_{T_h}=T_h\left(|d_{\pm1}\rangle \langle i_{\pm1}|+|d_{\pm2}\rangle \langle i_{\pm2}|\right)+h.c. 
\end{eqnarray*}
\begin{eqnarray*}
H_{T_h^f}=T_h^f\left(|d_{\pm1}\rangle \langle i_{\mp2}|+|d_{\pm2}\rangle \langle i_{\mp1}|\right)+h.c. 
\end{eqnarray*}
\begin{eqnarray}
H_{\mathrm exch}&=&\sum_{\substack{j=d,i}}\delta_0^j\left[ |j_{\pm1}\rangle \langle j_{\pm1}|-|j_{\pm2}\rangle \langle j_{\pm2}|\right]
\nonumber\\& &+\delta_1^j |j_{\pm1}\rangle \langle j_{\mp1}|+\delta_2^j|j_{\pm2}\rangle \langle j_{\mp2}|+h.c.
\end{eqnarray}
Here, $H_0$ describes the bare problem without interactions, with energies measured by $\delta_{d(i)}=E_{d(i)}-\hbar\omega_p$, as the exciton 
detuning with respect to the pumping laser energy, and where $\Delta_F=eFd$ is the Stark energy shift of the indirect excitons, where the electron
and hole are separated a distance $d$.  The interdot 
hole tunneling between states with the same total angular momentum and between bright and dark exciton states is described by 
$H_{T_h}$ and $H_{T_h^f}$, respectively. $H_{exch}$ accounts for the exciton fine structure splitting, where $2\delta_0^{d(i)}$ 
is the splitting between bright and dark exciton states, $\delta_1^{d(i)}$ characterizes the bright doublet mixing, and $\delta_2^{d(i)}$ 
that for the dark doublet.

The interaction with the incident laser is 
described by $H_p$, with optical coupling parameters $\Omega^d_{\pm}={\langle0|\vec{\mu}\cdot \vec{E}|d_{\pm1}\rangle}/{2\hbar}$ 
and $\Omega^i_{\pm}= {\langle0|\vec{\mu}\cdot \vec{E}|i_{\pm1}\rangle}/{2\hbar}$, associated to direct and indirect excitons, respectively.
Here $\vec{\mu}$ is the electric dipole moment which couples the excitonic transition to 
the electric component $\vec{E}$ of the radiation field. The schematic representation of the level configuration is shown in Fig.\ \ref{Fig1}, considering 
optical coupling, as well as hole tunneling--both conserving and non-conserving its angular momentum projection. 
We consider the condition $\Omega^d> \Omega^i$, as anticipated from the much lower e-h overlap in spatially indirect excitations. \cite{Rolon10} 
The electromagnetic field can be written as 
function of two orthogonal polarization components ($\sigma_+$ and $\sigma_-$), such that $\vec{E}=\vec{E}_{+}+\vec{E}_{-}$, 
where $E_+=E\cos(\phi)$ and $E_-=E\sin(\phi)$, and $0 \le \phi \le \pi/2$. 
Since the optical excitations obey selections rules conserving the angular momentum of $S_z=\pm1$, a circular right ($\sigma_+$, $\phi = 0$) 
or left ($\sigma_-$, $\phi = \pi/2$) 
polarized laser pulse creates intradot excitonic states $|d_{\pm1}\rangle$. 

To analyze the effects of the different couplings in the optical response of the QDM, it is convenient to define a new basis according to the  
symmetry of the states under parity: $|d_B^{S,A}\rangle\equiv\frac{1}{\sqrt{2}}\left(|d_{+1}\rangle\pm|d_{-1}\rangle\right)$,  $|i_B^{S,A}\rangle\equiv\frac{1}{\sqrt{2}}\left(|
i_{+1}\rangle\pm|i_{-1}\rangle\right)$, $|d_D^{S,A}\rangle\equiv\frac{1}{\sqrt{2}}\left(|d_{+2}\rangle\pm|d_{-2}\rangle\right)$ and $|i_D^{S,A}\rangle\equiv
\frac{1}{\sqrt{2}}\left(|i_{+2}\rangle\pm|i_{-2}\rangle\right)$, plus the exciton vacuum $|0\rangle$. The $+$ (or $-$) sign accompanies the symmetric $S$ 
(antisymmetric $A$)  superposition of exciton states, and the subindices $B$ and $D$ indicate whether the state of a given symmetry is a combination of 
bright or dark states, respectively. The advantage of this basis is that the two sets of symmetric and antisymmetric states are nearly disconnected, except
for their relative interaction with the vacuum through the radiation field.  This arises, as we show explicitly below, from the fact that the two processes of hole tunneling 
(spin preserving and not) couple direct and indirect excitons of only a given symmetry under parity. 

In the basis ordered by $|0\rangle$, symmetric ($|d_B^S\rangle$, $|i_B^S\rangle$, $|d_D^S\rangle$, $|i_D^S\rangle$) and antisymmetric combinations ($|
d_B^A\rangle$, $|i_B^A\rangle$, $|d_D^A\rangle$ and $|i_D^A\rangle$), we obtain the $9 \times 9$ hamiltonian $H^{'}$: 

\begin{widetext}
\begin{equation}
\label{eq:3}
H^{'}=\left(\begin{array}{ccccccccc}0&\Omega_d^S &\Omega_i^S &0&0&\Omega_d^A &\Omega_i^A &0&0\\
\cline{2-5}
\Omega_d^S &\multicolumn{1}{|c}{\delta_{d,B}^S}&T_h&0&\multicolumn{1}{c|}{T_h^f}&0&0&0&0\\
\Omega_i^S &\multicolumn{1}{|c}{T_h}&\delta_{i,B}^S&T_h^f&\multicolumn{1}{c|}{0}&0&0&0&0\\
0&\multicolumn{1}{|c}{0}&T_h^f&\delta_{d,D}^S&\multicolumn{1}{c|}{T_h}&0&0&0&0\\
0&\multicolumn{1}{|c}{T_h^f}&0&T_h&\multicolumn{1}{c|}{\delta_{i,D}^S}&0&0&0&0\\
\cline{2-5}
\cline{6-9}
\Omega_d^A &0&0&0&0&\multicolumn{1}{|c}{\delta_{d,B}^A}&T_h&0&\multicolumn{1}{c|}{-T_h^f}\\
\Omega_i^A &0&0&0&0&\multicolumn{1}{|c}{T_h}&\delta_{i,B}^A&-T_h^f&\multicolumn{1}{c|}{0}\\
0&0&0&0&0&\multicolumn{1}{|c}{0}&-T_h^f&\delta_{d,D}^A&\multicolumn{1}{c|}{T_h}\\
0&0&0&0&0&\multicolumn{1}{|c}{-T_h^f}&0&T_h&\multicolumn{1}{c|}{\delta_{i,D}^A}\\\cline{6-9}
\end{array}\right),
\end{equation}
\end{widetext}   
where $\Omega_d^{S(A)}=\frac{\Omega_+^{d} \pm \Omega_{-}^{d}}{\sqrt{2}}$, $\Omega_i^{S(A)}=\frac{\Omega_+^{i} \pm \Omega_{-}^{i}}{\sqrt{2}}$ 
describe the couplings between the ground state and symmetric (antisymmetric) combinations of bright exciton states. The diagonal terms related to the 
detuning of each level from the laser energy, as well as the splitting energy due to exchange interactions are given by: $\delta_{d,B}^{S(A)}=\delta_d+\delta_0^d 
\pm \delta_1^d$, $\delta_{i,B}^{S(A)}=(\delta_i-\Delta_F)+\delta_0^i \pm \delta_1^i$, $\delta_{d,D}^{S(A)}=\delta_d-\delta_0^d \pm \delta_2^d$, $\delta_{i,D}
^{S(A)}=(\delta_i-\Delta_F)-\delta_0^i \pm \delta_2^i$. The hamiltonian $H^{'}$ has a bordered block diagonal form, where the upper (lower) block is 
associated with the symmetric (antisymmetric) combination, and are coupled to the vacuum by the laser field. Notice that the anisotropic exchange interaction, represented in 
the model by $\delta_1^{d(i)}$, enters in the effective detuning of the bright symmetric (or antisymmetric) states, which couple to $|0\rangle$ by linearly polarized light given by 
the coupling $\Omega_{d(i)}^S$ ($\Omega_{d(i)}^A$). The $T_h$ and $T_h^f$ hopping amplitudes cause a mixture only between direct and indirect states of the same 
symmetry under parity. These features are evident on the exciton spectrum shown as function of the external electric field in Fig.\ \ref{Fig2}, where 
different symmetry states of the system mix within their class.

\section{Optical properties and exchange interaction}
\label{sec:chi}

The dynamical optical properties of the QDM are calculated using Liouville-von Neumann-Lindblad equations and we focus on the steady state of the system 
under continuous pumping. \cite{borges12} In order to characterize the optical response of the probe coupling, we calculate the optical susceptibility of an 
effective QDM ensemble with identical elements (which would of course characterize the response of a single QDM)
\begin{equation}
\label{eq:2}
\chi=\frac{\Gamma_{opt}}{V}\frac{|\mu_{0d}|^2}{\varepsilon_0\hbar\Omega_p}\left(\rho_{0,d_{\pm1}}+f \rho_{0,i_{\pm1}}\right),
\end{equation}
where $\rho_{0,d_{\pm1}}$ and $\rho_{0,i_{\pm1}}$ are the density matrix elements associated with allowed optical transitions, $\Gamma_{opt}$ is the 
optical confinement factor given by the number of QDMs in a given region of the ensemble, $V$ is the volume of a single QDM, $
\varepsilon_0$ is the host dielectric constant and $f={\mu_{0i}}/{\mu_{0d}}\ll 1$ is the ratio between the interband dipole moments of the indirect and direct 
transitions. The susceptibility is a complex function, written as $\chi={\chi}'+i{\chi}''$, where the absorption coefficient $\alpha(\omega_p)$ is proportional to 
${\chi}''$, while the refractive index $n(\omega_p)$ depends on both, real and imaginary parts of $\chi$. \cite{Koch}

In the following we consider realistic parameters for InAs self-assembled QDMs and exciton energies in agreement with experimental measurements. \cite{Stinaff11} 
Exchange interaction in these structures typically results in splittings between dark and bright states of 
$\delta_0^d \simeq 100~\mu\mathrm{eV}$, $\delta_1^{d}
\simeq 35~\mu\mathrm{eV}$ and $\delta_2^{d} \simeq 10~\mu\mathrm{eV}$. \cite{Bracker08}  As the overlap of the electron-hole wave functions 
is smaller in indirect 
states, we also assume $\delta^i_{0,1,2}\approx 0.1\delta^d_{0,1,2}$. We also consider effective relaxation parameters $\Gamma_d=10\mu\mathrm{eV}$, 
$\Gamma_i=10^{-3}\mu\mathrm{eV}$, typical in experiments.\cite{Borri03} The tunneling coupling, which depends strongly on the barrier 
is assumed comparable to the decay 
rate of the direct state, $T_h = \Gamma_d/2$ and $T_h^f=T_h/2$. From the conditions $\Omega^d\approx 0.1\Gamma_d$ and $\Omega^i\approx 
0.1\Omega^d$, it follows that $\Omega_d, \Omega_i\ll T_h,T_h^f$.  Other parameters in the susceptibility, including the optical confinement factor 
$\Gamma_{opt}=6 \times 10^{-3}$, momentum matrix element $\mu_{0d}=21${\AA}, and QDM volume $V\approx 800${\AA}$^3$, set the overall 
strength of the optical response, and were taken from Kim {\em et al}. \cite{Kim04}

\begin{figure}[h]
\centering
\includegraphics[scale=1.0]{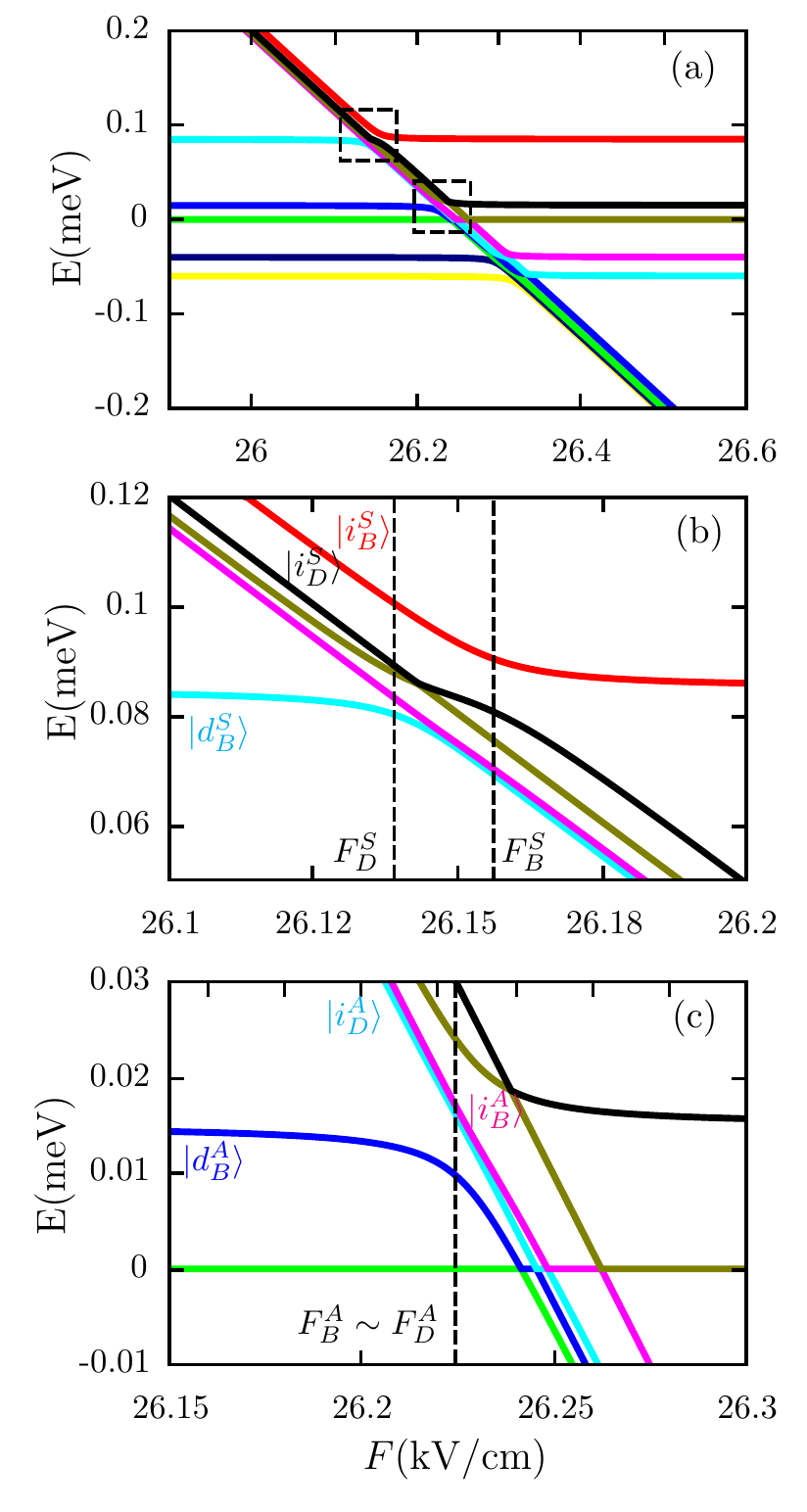}
\caption{(Color online) (a) Exciton spectrum as function of electric field $F$. Boxed regions show the relevant direct and indirect excitons anticrossing. (b) 
and (c) show an amplification of the regions where the 
direct excitons $|d_{B}^S\rangle$ and $|d_{B}^A\rangle$ anticross with indirect exciton states of the same symmetry under parity, S and A, respectively. 
Vertical dashed lines indicate electric field 
$F$ values where the resonance condition between direct and indirect excitons is fulfilled.}\label{Fig2}
\end{figure}
 
The excitonic spectrum exhibits characteristic splittings due to the exchange interaction and anticrossing regions which are signatures of the mixture between 
direct and indirect exciton states, as shown in Fig.~\ref{Fig2}(a).  Let us concentrate in the anticrossing regions indicated by dashed squares in the figure.
Figures \ref{Fig2}(b) and \ref{Fig2}(c) show a zoom of the highlighted areas in \ref{Fig2}(a), illustrating that $T_h$ and $T_h^f$ couplings lead to 
the expected anticrossings. In Fig.~\ref{Fig2}(b) we can see clearly that $|d_B^S\rangle$ is coupled only to two indirect symmetric states: 
$\vert i_B^S\rangle$ and $\vert i^S_D \rangle$. The mixing with the $|i_B^S\rangle$ state occurs due to $T_h$, and with $|i_D^S\rangle$ through $T_h^f$. 
A similar anticrossing structure can also be observed in Fig.~\ref{Fig2}(c) for asymmetric states, where the $\vert d_B^A\rangle$ state couples with $\vert 
i_B^A\rangle$ ($\vert i_D^A\rangle$) due to $T_h$ ($T_h^f$). 
The electric field value at which the anticrossing occurs is obtained by the resonance condition between states $|d_B^S\rangle$ and $|i_B^S\rangle$, $
\delta_d+\delta_0^d+\delta_1^d=(\delta_i-\Delta_F)+\delta_0^i+\delta_1^i$, which yields $F_B^S=\Delta_F /ed$, whereas $F_D^S$ is the 
field value at which the states $|d_B^S\rangle$ and $|i_D^S\rangle$ meet and anticross, as indicated in 
Fig.~\ref{Fig2}(b). 

Figure \ref{Fig2}(c) shows the region of anticrossing between the state 
$|d_B^A\rangle$ and the indirect states $|i_B^A\rangle$ (for $F=F_B^A$) and $|i_D^A\rangle$ (for $F=F_D^A$), with field values also 
obtained through the resonance conditions between the corresponding states (notice that $F_B^A \simeq F_D^A$ for our choice of parameters).

\begin{figure}[b]
\centering
\includegraphics[scale=0.67]{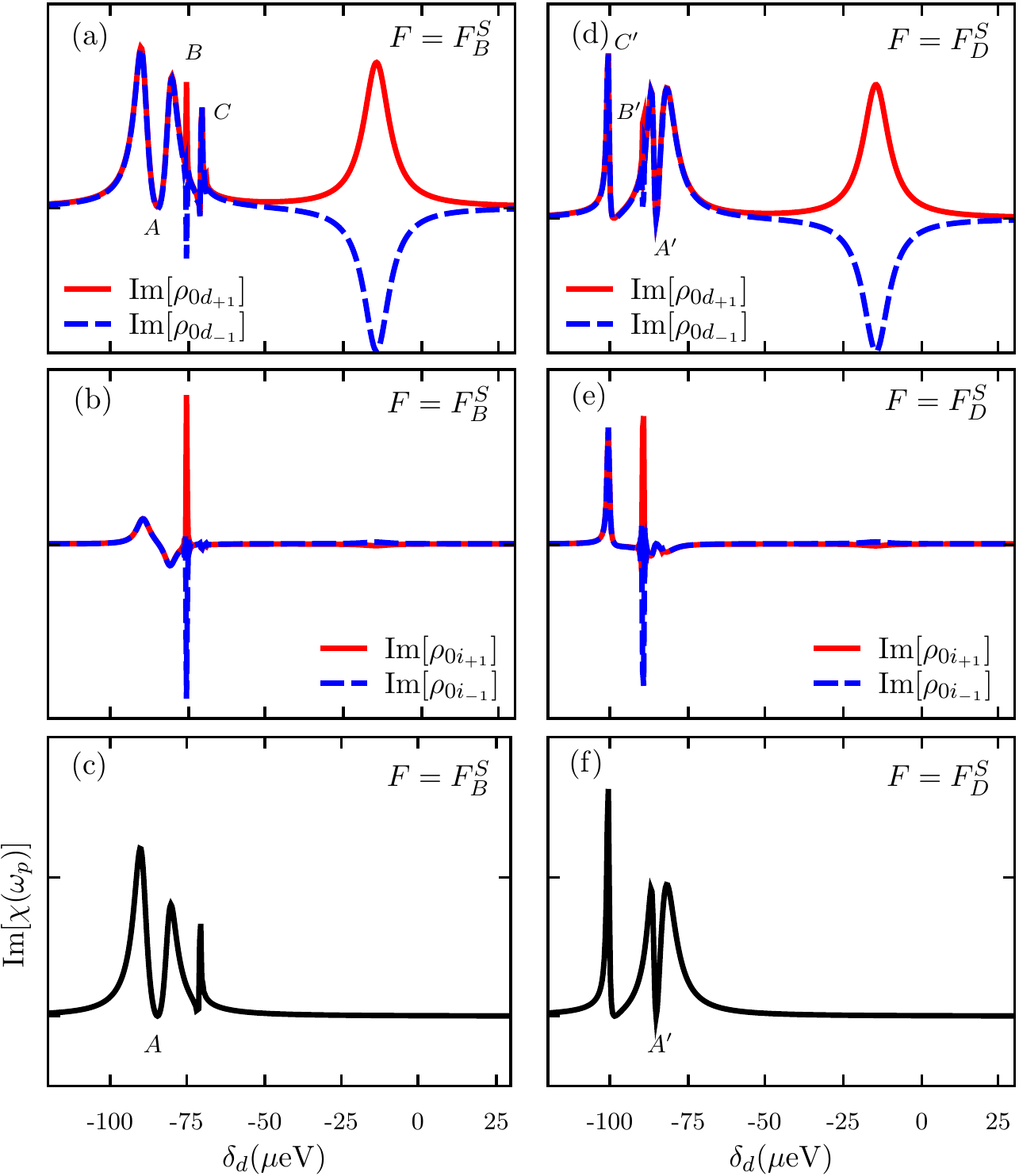}
\caption{(Color online) Imaginary part of the density matrix elements contributing to $\chi(\omega_p)$ under right circularly polarized laser light ($\phi=0$). 
We show in (a) and (d) the direct exciton contribution $\rho_{0,d_{\pm 1}}$; (b) and (e) show the indirect exciton contribution $\rho_{0,i_{\pm 1}}$; (c) and (f) 
show the total imaginary part of the optical susceptibility, $\mathrm{Im}\chi$, which describes absorption by the QDM. 
Left panels are for $F=F_D^S$ and right panels for $F=F_D^S$. $A$ and $A'$ labels indicate the window of transparency induced by the
tunneling couplings $T_h$ and $T_h^f$, respectively.}\label{Fig3}
\end{figure}

We now investigate the conditions for which the couplings $T_h$ or $T_h^f$ can establish efficient destructive quantum interference paths and produce a 
subsequent reduction on the absorption spectrum. To this end we analyze the optical susceptibility $\chi$ for electric field values 
where the anticrossings are observed. It is important to recall that although the total optical absorption of the system (\ref{eq:2}) has contributions of different weights 
related with direct ($\rho_{0,d_{\pm 1}}$) and indirect states ($\rho_{0,i_{\pm 1}}$), the oscillator strength of spatially separated carriers is much smaller 
than for the direct exciton, and $\rho_{0,i_{\pm 1}}$ does not contribute significantly, as we will see.

Figure \ref{Fig3}(a) and (b) show the imaginary part of the density matrix elements $\rho_{0,d_{\pm 1}}$ and $\rho_{0,i_{\pm 1}}$ 
respectively, as function of the detuning $\delta_d$ for field $F_B^S$ where bright symmetric states $|d_B^S\rangle$ and $|i_B^S\rangle$ anticross, 
considering right circularly polarized laser $\sigma_+$ ($\phi=0$).  Panel \ref{Fig3}(c) shows the total absorption profile given by $\mathrm{Im}
[\chi(\omega_p)]$ as a function of $\delta_d$, taking into account all contributions.  It is interesting to see how the different features in these curves
arise.

Figure \ref{Fig3}(a) shows the separate contributions of both matrix elements $\rho_{0d_{+1}}$ (solid red line) and $\rho_{0d_{-1}}$ (dashed blue line). We observe two 
regions of absorption, the first one around $\delta_d=-\delta_0^d+\delta_1^d=-15\mu\mathrm{eV}$, corresponds to the resonance of the probe laser with 
the asymmetric state 
$|d_B^A\rangle$, while the second one at $\delta_d=-\delta_0^d-\delta_1^d= -85\mu\mathrm{eV}$ corresponds to resonance 
with state $|d_B^S\rangle$. We observe that at $\delta_d=-15\mu\mathrm{eV}$, the asymmetric contributions to absorption $\rho_{0,d\pm 1}$ cancel 
each other. In contrast, when the probe laser is resonant to the symmetric transition, the two contributions are additive, and a dip (labelled as \emph{A} in 
Fig.~\ref{Fig3}(a)) is observed at $\delta_d = -85\mu\mathrm{eV}$. This dip in the absorption is due to the hole hopping amplitude $T_h$ that couples the states $|d_B^S
\rangle$ and $|i_B^S\rangle$; the mixing provides a destructive interference path creating the narrow transparency window labelled \emph{A}. 
 In addition, Fig.~\ref{Fig3}(a) shows two lateral peaks, \emph{B} and \emph{C} that occur at smaller detuning and are due to the coupling of state $|d_B^S\rangle$ with other 
 quasi-resonant states when $F=F_B^S$. The lateral peak \emph{C}, at $\delta_d\approx -70\mu\mathrm{eV}$ is especially an additive contribution of 
 $\rho_{0d_{\pm1}}$ due to the $T_h^f$ coupling between $|d_B^S\rangle$ and $|i_D^S\rangle$, and provides an optical signature of the presence of
 ``spin-flipping" events in the interdot tunneling.
 
\begin{figure}[htb]
\centering
\includegraphics[scale=0.67]{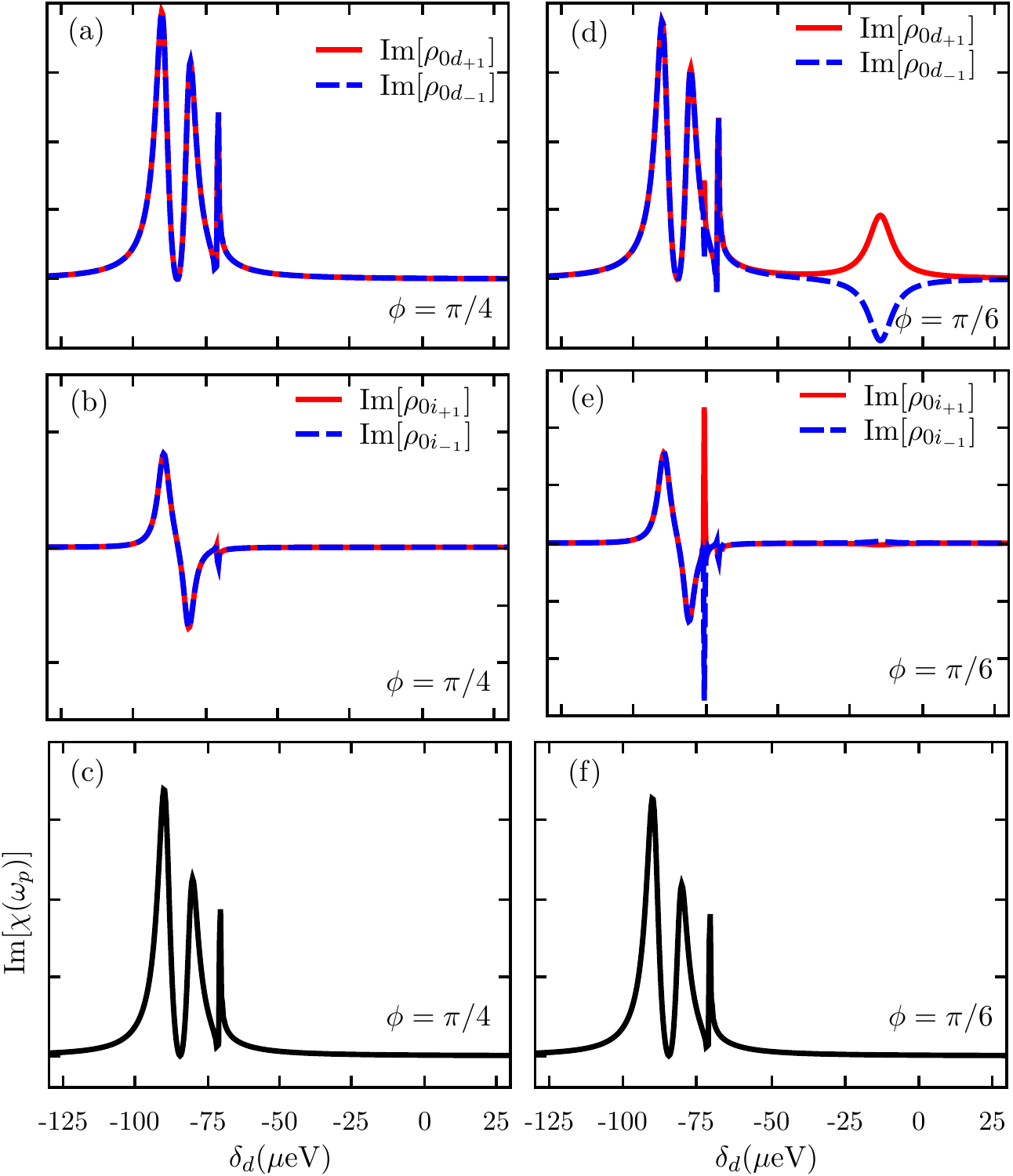}
\caption{(Color online) Imaginary part of the density matrix elements that contribute to total optical 
susceptibility $\chi(\omega_p)$ as a function of $\delta_d$ for different probe polarization conditions. Left panels are for $\phi=\pi/4$ (linear polarization) and 
right panels for $\phi=\pi/6$ (elliptical polarization). In all situations $F=F_B^S$.}\label{Fig4}
\end{figure}

The contributions labeled \emph{B} in Fig.~\ref{Fig3}(a) also modify the 
optical response of the QDM in an interesting way, as can be seen in Fig.~\ref{Fig3}(c).  This deserves a more detailed analysis. The optical coupling 
of the ground state with 
direct and indirect excitons induces an effective coupling between states of different symmetry, as seen in the hamiltonian~(\ref{eq:3}). This effective 
coupling can be seen to arise from second-order terms that can be obtained by tracing out (projecting out) the $|0\rangle$ state of the total 
hamiltonian. \cite{Cohen} 
Considering, for instance, only the bright direct exciton states, the resulting effective Hamitonian (ground state projected out) written in the basis $|d_B^S
\rangle$ and $|d_B^A\rangle$ is given by:
\begin{equation}
\label{eq:4}
H_{\mathrm eff}^{d,B}=\left(\begin{array}{cc}\delta_{d,B}^S+\frac{\left(\Omega_d^S\right)^2}{\delta_{d,B}^S}&\frac{\Omega_d^S\Omega_d^A}{2}\left(\frac{1}
{\delta_{d,B}^S}+\frac{1}{\delta_{d,B}^A}\right)\\
\frac{\Omega_d^S\Omega_d^A}{2}\left(\frac{1}{\delta_{d,B}^S}+\frac{1}{\delta_{d,B}^A}\right)&\delta_{d,B}^A+\frac{\left(\Omega_D^A\right)^2}{\delta_{d,B}
^A}
\end{array}\right),
\end{equation}
with optical couplings $\Omega_d^{S(A)}$, and detuning parameters $\delta_{dB}^{S(A)}$ as defined in Sec.\ \ref{sec:theory}.
It is clear from (\ref{eq:4}) that there is an effective coupling between 
symmetric $|d_B^S\rangle$ and antisymmetric $|d_B^A\rangle$ direct exciton states, which although weak ($\approx \Omega_d^S \Omega_d^A$), it is not zero. 
The effect on the direct exciton density matrix elements are peaks identified 
as \emph{B} in Fig.~\ref{Fig3}(a). 
The effect on the optical response is nearly null, however, because \emph{B} peaks on Im[$\rho_{0d\pm 1}$] nearly cancel each other out, as shown in Fig.~\ref{Fig3}(a) for $\delta_d 
\approx -75 \mu \mathrm{eV}$. Only the peak \emph{C}, associated to the mixing between $|d_B^A\rangle$ and $|i_D^A\rangle$ and due to $T_h^f$, 
contributes significantly to the net absorption, as seen in Fig.~\ref{Fig3}(c), at $\delta_d=-70 \mu \mathrm{eV}$.

A similar but somewhat complementary behavior on the absorption spectrum can be obtained at the field $F=F_D^S$ where the resonance condition $
\delta_d+\delta_0^d+\delta_1^d=(\delta_i-\Delta_F)-\delta_0^i+\delta_2^i$ is satisfied and the $|d_B^S\rangle$ and $|i_D^S\rangle$ states mix.
As before, Figs.~\ref{Fig3}(d) and (e) show the contribution of the four matrix elements associated to the total absorption. We observe this time a transparency 
window on the main absorption due to $T_h^f$ (labeled as \emph{A}$^\prime$) at $\delta_d \approx -85 \mu \mathrm{eV}$, and is therefore narrower than the 
dip \emph{A} in Fig.~\ref{Fig3}(c). 
The width of the transparency window is directly proportional to the strength of the coupling that induces the destructive interference process, which in this 
model is given by $T_h$ or $T_h^f$. As we have assumed (and expect) $T_h^f<T_h$, the difference on the transparency windows is evident in the dip induced by the different 
couplings. 
In this case, the presence of the lateral peak \emph{C}$^\prime$ observed in the total absorption spectrum at more negative detuning (see Fig.~\ref{Fig3}(f)), is due to $T_h$ which 
couples $|d_B^S\rangle$ and $|i_B^S\rangle$. In both cases we observe strong asymmetry on the height of the peaks induced by the contribution of indirect 
states in the optical coupling. 

The level anticrossing of the asymmetric manifold that occurs at $F^A_D \simeq F^A_B$ (see Fig.\ \ref{Fig2}(c)) also results in structure for the different 
density matrix elements.  However, as in other anticrossings, the contributions to $\rho_{0d_{\pm1}}$ and $\rho_{0i_{\pm1}}$ are out of phase and do not 
contribute to the overall absorption structure of the system.  In other words, the resulting absorption function ($\mathrm{Im} \chi$) at or near those field values 
shows no appreciable tunneling induced transparency window, but rather a single absorption peak on resonance (not shown).

Another parameter that can be easily controlled experimentally is the laser polarization. Figure \ref{Fig4} shows the total absorption considering the four 
contributions of the density matrix elements for two different $\phi$ values, $\phi=\pi/4$ (linear polarization) and $\phi=\pi/6$ (elliptical polarization), for 
$F=F_B^S$.
In full agreement with the hamiltonian (\ref{eq:3}), in Fig.~\ref{Fig4}(a) we note that a linearly polarized probe laser couples only the states $|0\rangle$ 
and $|d^S_B\rangle$. Therefore we observe in this figure only the features associated to the symmetric state.  When the incident laser is polarized elliptically, 
Fig.~\ref{Fig4}(d), the optical couplings are different, $\Omega_+^{d(i)}>\Omega_-^{d(i)}$, and the amplitude of the peak associated to $|d^A_B\rangle
$ in the imaginary part of $\rho_{0,d_{\pm}}$ is smaller. However, the total absorption spectrum exhibits the same behavior for different polarization 
conditions, Figs.~\ref{Fig4}(c) and (f), since the main difference occurs in the peaks associated to the asymmetric state for $\delta_d=-15\mu\mathrm{eV}$. 

\begin{figure}[t]
\centering
\includegraphics[scale=0.7]{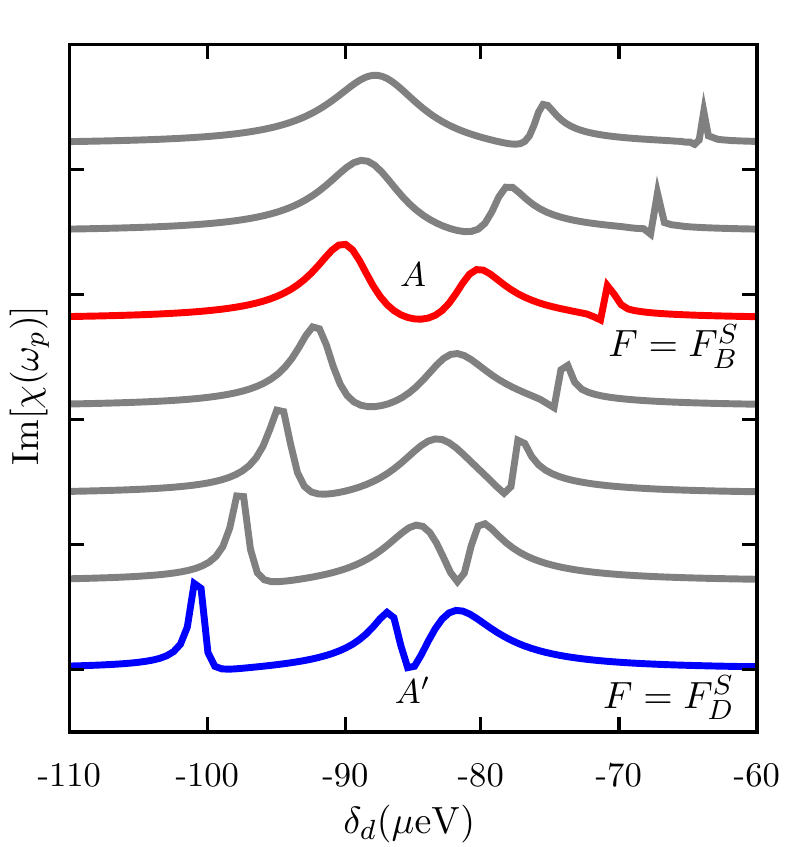}
\caption{(Color online) The imaginary part of $\chi(\omega_p)$, as function of detuning $\delta_d$ for different field values separated by $4.375~
\mathrm{V/cm}$, shows the evolution of the transparency windows $A$ and $A'$.  Curves offset for clarity.}\label{Fig5}
\end{figure}

As the presence of the lateral peak (\emph{C} and \emph{C}$^\prime$) and the transparency window 
(\emph{A} and \emph{A}$^\prime$) are directly related to the tunneling couplings $T_h$ and $T_h^f$, we plot in Fig.~\ref{Fig5} the total absorption as 
function of the detuning $\delta_d$ for different values of electric field, over a range of $26~\mathrm{V/cm}$ that includes the field values  $F_B^S$ and $F_D^S$
where the anticrossings occur. Notice the much expanded horizontal scale from that in Fig.~\ref{Fig3}(c) and (f), to better appreciate details. The lines 
at fields where the anticrossing occur are indicated in the figure. The $A$ and $A'$ absorption dips due to destructive interference by tunneling are also 
indicated.  It is interesting to notice how they slowly evolve away from the resonant field to create additional lateral absorption features 
whenever they are away from resonance.

\section{Conclusions}
\label{sec:concl}
In summary our results show that through the appropriate control of external electric field and light polarization,
the tunneling coupling between states of the same total angular momentum ($T_h$) and between bright and dark states states ($T_h^f$) establishes 
efficient destructive quantum interference paths, creating transparency windows in the absorption spectra with controllable strength and position. This 
effect could be inversely used to characterize the strength of different exchange couplings in QDM, as the polarization and Stark shifts can be 
independently controlled while the absorption spectrum of the system is monitored. Similar coherent superposition of states and the appearance of interfering paths
could be explored in other excitonic manifolds in these QDMs, including biexcitons and charged excitons, as they may provide the ability to control the polarization of
the outgoing photons as fields and laser detunings are carefully adjusted.

\acknowledgments

We thank helpful discussions with J. M. Villas-B\^oas, 
E. Stinaff, and R. Thota, as well as the support of CAPES, CNPq, Fapemig, and INCT-IQ, Brazil, as well as US NSF CIAM/MWN grant DMR-1108285.

\bibliography{eit3}
\end{document}